\documentclass[1opt]{article}
\usepackage{graphicx,amsmath,amsfonts,amssymb,dcolumn,amsthm,euscript}
\def\bea{\begin{eqnarray}}
\def\eea{\end{eqnarray}}
\def\be{\begin{equation}}
\def\ee{\end{equation}}
\usepackage{color}

\newcommand\nn{\nonumber} 
\newcommand{\bq}{\begin{equation}}
\newcommand\eq{\end{equation}}

\def\bar{\overline}
\newcommand\pa{\partial}

\def\d{{\rm d}}

\newcommand\del{\delta}
\newcommand\lam{\lambda}
\newcommand\Ph{\varPhi}
\newcommand\Ps{\varPsi}

\usepackage{slashed}

\def\demi{\frac{1}{2}}

\begin{document}
\title{\textbf{ 
 {   
Unimodular Gauge and   ADM Gravity Path Integral  
}   \\$ $\\ 
 }}
\author{\textbf{Laurent Baulieu
 }
\thanks{{\tt baulieu@lpthe.jussieu.fr}
}
\\\\
\textit{
LPTHE, Sorbonne Universit\'e, CNRS} \\
\textit{{} 4 Place Jussieu, 75005 Paris, France}\\
}

\def\blue{  \color{blue}}
\def\red{  \color{red}}
\def\green{  \color{green}   }
\def\yelllow{  \color{yellow}}

\date{  
}
\maketitle
\begin{abstract}
This paper proposes a definition of gravitational observables and of their path integral formula within the framework of ADM foliation and the choice of unimodular gauge classes. The method enforces a BRST invariant gauge fixing of the lapse and shift fields. It yields the quantum level extension of the known classical property that the conformal classes of internal metrics of constant Lorentz time leafs define the gravitational physical degrees of freedom.

\end{abstract}

\def\blue{  \color{blue}}
\def\red{  \color{red}}
\def\yelllow{  \color{yellow}}
\def\black{  \color{black}   }

  $ $

\def\s{{s_{stoc}}}
\def\hl{{\hat l}}

\def\l{\lambda}
\def\ll{{\hat \lambda}}
\def\th   {\theta}
\def\thh{{\hat \theta}}
\def\ee{{\epsilon}}
\def\eeh{{\hat \epsilon}}
\def\dd{\delta}
\def\ddh {\hat \delta}

\def \pa{\partial}
 \def\l{{\lambda}}
  \def\g{{\gamma}}
   \def\T{{t}}
   \def\t{\tau}
  \def \d{\delta}
 \def\d{{\rm d}}
 \def\x{{q}}

 \def   \hg { {\hat g}}
  \def\g {{\sqrt{g}}}
  
  \def\bx  {
   {  {{\bar \xi}}^{(01)   } } }
   
  \def\bxm {
   {  {{\bar \xi}}^{(01)  \mu} } }
   
    \def\bxn {
   {  {{\bar \xi}}^{(01)  \nu} } }
   
     \def\L  {
   {  {{L }}^{(00)   } } }
   
   \def\Lx{     Lie_{\xi   }  }  
  
      \def\s{  \hat  s   }
          
    \def\hg{    {\hat g}} 
    
       \def\g{    {\sqrt  {g}  }}
 
 \def\v{ {\varphi}}
     \def\s{     s   }
     \def \vx {\vec{x}}
     \def\V{   {{\Phi}    }}
     \def \Vx { \V (\vx)}
     \def\vv{   {\vec {v}    }} 
     \def\vV{   {\vec {V}    }}
  \def \vX {\vec{X}}

  \def\blue{  \color{blue}}
\def\blue{  \color{black}}

\def\red{  \color{red}}

     \def\bgij{\bar g  _{ij}}

\def\s{{s_{stoc}}}
\def\hl{{\hat l}}

\def\l{\lambda}
\def\ll{{\hat \lambda}}
\def\th    {\theta}
\def\thh{{\hat \theta}}
\def\ee{{\epsilon}}
\def\eeh{{\hat \epsilon}}
\def\dd{\delta}
\def\ddh {\hat \delta}

\def \pa{\partial}
 \def\l{{\lambda}}
  \def\g{{\gamma}}
   \def\T{{t}}
   \def\t{\tau}
  \def \d{\delta}
 \def\d{{\rm d}}
 \def\x{{q}}

 \def\gmn{g_{\mu\nu}}

 \def   \hg { {\hat g}}
  \def\g {{\sqrt{g}}}
  
  \def\bx  {
   {  {{\bar \xi}}^{(01)   } } }
   
  \def\bxm {
   {  {{\bar \xi}}^{(01)  \mu} } }
   
    \def\bxn {
   {  {{\bar \xi}}^{(01)  \nu} } }
   
     \def\L  {
   {  {{L }}^{(00)   } } }
   
   \def\Lx{     Lie_{\xi   }  }  
  
      \def\s{  \hat  s   }
          
    \def\hg{    {\hat g}} 
    
       \def\g{    {\sqrt  {g}  }}
 
 \def\v{ {\varphi}}
     \def\s{     s   }
     \def \vx {\vec{x}}
     \def\V{   {{\Phi}    }}
     \def \Vx { \V (\vx)}
     \def\vv{   {\vec {v}    }} 
     \def\vV{   {\vec {V}    }}
  \def \vX {\vec{X}}

  \def\vp{\varphi}

\def\demi{\frac{1}{2}}
\def\pa{\partial}
\def\d{{\mathrm d}}
\def\tpa{\tilde\pa_\tau} 
\def\k{D_\tau{\ln\sqrt g}}
\def\Diff{\mathrm{Diff}_\Sig ({\cal M})}
\def\Weyl{\mathrm{Weyl}_{\cal M}}
\def\Lie{\mathrm{Lie}}
\def\gauge{\mathrm{gauge}}

\def\Ph{\varPhi}
\def\Ps{\varPsi}
\def\Del{\varDelta}
\def\Gam{\varGamma}
\def\Sig{\mathsf{\Sigma}}
\def\del{\delta}
\def\lam{\lambda}
\def\fp{\varphi}
\def\A{{\pmb a}}
\def\NF{{\cal N}}

\def\nn{\nonumber}

  \def\gij{{g_{ij}}}
\def\hgij{{ \bar g_{ij}}}
\def\hgmn{{ {\hat g}_{\mu\nu}}}
  \def \A{\alpha}
  
   \def\N{{N^{(i)}}}
    \def\N{{N^i }}
   \def\dg{\sqrt {\det{  \gij}}}

 \newpage
\black
 \section{Introduction}
 
 \def\bluecor{\color{blue}}
  \def\bluecor{\color{black}}
 The purpose of this note is   to enlighten   the definition of gravitational observables and  of   the quantum gravity path  integral. To do so,  one  uses  the choice of   unimodular gauge classes within     the  framework of the 
 ADM foliation   \cite { Arnowitt:1962hi}.

Giving a central  role to   the unimodular gauge  is  inspired by the
observation that what really matters in classical gravity is the propagation
of conformal classes of spatial metrics. 
To the best of our knowledge, this classical  and precise property was firstly advocated in the
physical literature in~\cite{York} where part of Einstein
equations of motion is  cornered out  as  describing only physically
irrelevant propagation of constraints.~\cite{York} shows  that    
  the truly relevant initial physical data for solving Einstein equations
of motion  are the  conformal classes of spatial metrics. {In fact, mathematicians found already in $1925$ the
relevance of Weyl symmetry for solving Einstein equations
\cite{Thomas199, Thomas2, Thomas3}, a property that Einstein has foreseen even  earlier \cite{Einstein}.  Using  the Weyl invariance in the quantization of gravity   turns out to be very natural in the stochastic quantisation framework \cite{bcw} and 
    a BRST invariant     unimodular gauge fixing has been  recently defined     such  that  the conformal factor 
    can be gauge fixed for having   no physical propagation, while the theory  maintains a BRST invariance \cite{unimodular}.  However~\cite{unimodular}  is not using the ADM paradigm. This  is perfectly fine and    of interest     for   some  specific questions (such as the perturbative propagation of gravitational quanta).   However, the    ADM parametrization is   most relevant    in some cosmological problems due to  it  visualisation   of the space time  as a  space being foliated by the Lorentz time rather than as a  more disorganized  set of points. Therefore it makes sense to address the question of  how to formulate the unimodular gauge in the  ADM foliation framework.

  {
  \bluecor  Both  terms   ``unimodular gauge"  and      ``unimodular gravity" should not be confused. The point is that  Einstein 
 recommended  as early as in 1916  that one  uses   systems  of coordinates  to    solve his  gravity  equation    such that the determinant  of the  space-time metric $g_{\mu\nu}$~is    locally unimodular~\cite{Einstein}. This     has been the source of  many subsequent works and    the word   ``unimodular gravity"~has   become quite common.    But the  so-called ``unimodular gravity" has    a  different  physical content than the  Einstein theory   gauge  fixed in a standard BRST invariant way (in particular in the case  of the unimodular gauge). The  non-exhaustive  series of papers 
\cite{Henneaux}\cite{Alvarez1}\cite{Padilla}\cite{Bufalo}\cite{Alvarez2}\cite{Oda1}\cite{Oda2}\cite{Percacci}\cite{Nagy}\cite{Martin}\cite{Vikman}
and  references therein    address      questions     related to this domain.  When looking at the literature,
  there are basically  two formulations: one  that imposes $g= -\det(g_{\mu\nu})=1$ as a gauge choice and another one  that    imposes the constraint $  g=1 $ from the beginning.  Some confusion  is   spread  around   these formulations, although their difference is  quite  clear.  
  
Working  out the Einstein theory    in the    class of   the ``unimodular gauges"   where the  metric determinant  $g$ is  fixed     in a BRST invariant way 
is   nothing but    a possible gauge choice, perhaps unfamiliar  and difficult to   enforce, but formally equivalent to any  other   (consistent) gauge choice, as can be proven when one uses consistently the  BRST~methodology to enforce  both 
gauge choices.    In fact,  
the present paper  pushes further  the BRST  framework to     explains   how  the unimodular gauge choice  can be done within 
the ADM metric parametrization. One motivation is that  the combination of using unimodular gauge and the 
ADM paradigm  makes more transparent the extension  at a   path integral  level  of  the classical  property    the gravity observables  can be represented as functionals of the  unimodular part of the metric.
This latter point relies on    
the  physical  degeneracy  between  the ADM space like leafs  related by a Weyl transformation when they are used as an initial condition for solving the Cauchy problem expressed by the Einstein equations, as   explained  long time ago  by York~\cite{York}. Getting a proper definition of the observables is   an important component of the BRST paradigm.  It is a   quite  sophisticated  concept  in gravity since its local reparametrization symmetry associated to  the propagation of the metric cannot be treated stricto sensu as a gauge symmetry and certainly not on the same footing as  its   local Lorentz symmetry\footnote{ The involved  subtleties share    analogies  with those occurring in the definition of observables in  topological quantum field theories.}. But a  manifestly BRST invariant method ensures  that observables, once they are well defined at the classical level,  are independent of gauge artefacts at the quantum level.. 

 In contrast,      ``unimodular gravity"   means  changing  the theory by varying classically  the 
Einstein--Hilbert action with nothing  but    variation of   metrics with $\sqrt g=1$. A given motivation of 
  ``unimodular gravity"  has been  that it  introduces the cosmological constant    as a   constant of integration to    be chosen at will 
while  the  ``unimodular gauge"   fixes the cosmological constant as a parameter of the Lagrangian   from the beginning.
In fact, 
 the   ``unimodular gravity"      path integral  is supposed to be a genuine     summation  over unimodular metric. 
 It  postulates that  the symmetry group is reduced from   Diff(M$_d$)  to ``transverse"  Diff(M$_d$).  This restriction    makes sense at the 
classical level, but   its   brute force enforcement at the    quantum level  does't come with  a corresponding Faddeev-Popov ghosts and ghost of ghosts compensating ghost action. 

 It is yet  quite    unclear if the ``unimodular gravity"   approach is  a  consistent one.  Indeed,  it comes in contradiction  with the by now rather well accepted  principle that: 
(i)~the   local quantum field theory  one may associate to   given a classical  theory with  a  local symmetry  must be built while taking into account the  BRST symmetry associated to the whole classical local symmetry;  (ii)~its gauge-fixing must be done by building   a  ghost number zero BRST invariant    action that is  a local functional of  all   interacting geometrical ghosts  (and possibly ghosts of ghosts)   with no ghost zero modes  as well as of   all the trivial BRST doublets of antighosts and Lagrange multipliers that are adapted to the desired gauge choice  - in the case of gravity the ghosts and antighosts  must be introduced  to consistently define the BRST symmetry  based on the  complete  Diff(M$_d$) and  in the case of the  unimodular gauge the ghost and antighost structure is more elaborated than for instance in the deDonder gauge -;   (iii) the definition of
the physical observables must be done in a way that is compatible with  both  requirement of BRST symmetry and 
of unitarity\footnote
{\bluecor
    More must be said about the stability of a  given choice of a classical gauge function.   In   Yang--Mills, the currently used linear gauges are stable because  one can complete  the gauge symmetry BRST Ward identities  by additional ones that are consequences of   the antighost equations of motion.  The latter ensure  the stability of linear  gauge fixing.   But the non linearity of gravity   makes it more difficult to prove that     the unimodular gauge fixing   is   stable under radiative corrections  when one  introduced order by  order in perturbation theory the   needed counterterms that are compatible with the Ward identities in this gauge, although its physicality is a good argument for     this stability to be true. This question deserves further   investigations.}.

 In fact,    the ways one formulates     both  
``unimodular gauge gravity" and    ``unimodular gravity"  are so different that  they  have a very good chance to   provide  non equivalent
  quantum versions.
 Higher order perturbative computations of quantum gravity might be the clue to reveal their  differences, but, in  our knowledge, no  such computations have  yet appeared in the literature.

 
 }
 {

 The logics  of  "unimodular gauge gravity"  was explained in  \cite{unimodular} in the context of the BRST quantization method when the fundamental fields    that appear in the gravitational path integral mesure  are   the genuine~$d$  metric  components $\gmn$. \cite{unimodular} 
 relies   on the correct property that the condition  $\sqrt g=1$  is a   well-defined classical local gauge condition for gravity that   is allowed by    the reparametrization  invariance  of the classical  theory. As a matter of principle,   no ambiguity   exists in this gauge  to define  a semi classical     quantum field theory   of   gravity, provided that it    is consistently enforced in a BRST invariant way  in the path integral,  as sell as  the remaining of the reparametrization  is also consistently gauge fixed  for the unimodular part   $\hat g_{\mu\nu}$ of the $d$  metric. \cite{unimodular} 
   solved the question of building a ghost antighost system  such that no zero modes  is  created in the path integral. 

The   present work   extends the proof  of   \cite{unimodular} 
 within the ADM paradigm. It 
   builds   a class of  BRST exact    actions           that  enforces  consistently the   unimodular gauge condition. The latter, which are local  functions of the ADM fields and their ghosts, must  be    
   added  to  the  classical ADM Einstein action.     This provides   a definition of  the gravity  path integral within  the ADM formalism in the ``unimodular gauge gravity"  that is a consistent  one (at least semi classically).   
    No  possible anomaly is expected  in  this process but the standard   gravitational anomalies that can   possibly exist from a purely geometrical point view, due  to  the structure of the $SO(d-1,1)$ Lie algebra. 

   }


      The   Lagrangian  foliation  framework      separates   the handling of  the gravitational  constraints
from the   quantum    gauge-fixing   of the $d$ dimensional~reparametrization symmetry.  
 At the classical level,  the ADM lapse and shift   have indeed no dynamics.  They are    basically  spectators.    
 At the quantum level, 
 the BRST invariant gauge fixing for the unimodular gauge in the ADM framework        is quite relevant to clarify how this property extends at the quantum level.  It       justifies  quite transparently the    definition of the quantum gravity  observables as   functionals of the unimodular parts of the  Euclidean $d-1$ dimensional  leaf internal metric, as   fundamentally   inspired    by York  classical gravity results~\cite{York}.
  
  Thus,    calling  $\bar g_{ij}$ the  unimodular part
of 
the spacelike    $d-1$  dimensional  internal metrics    $ g_{ij}$  of the  ADM~leafs that foliate    with respect to the  Lorentz time
the $d$ dimensional   space with  pseudo Riemannian  $d$    metric   $\gmn$, this paper  suggests that the physical content  of  quantum gravity  is described by the  expectation values 
 \bea\label{?}
 <{\cal O} [\bar g_{ij}]>=\rm {as\  it  \ will \ be \ defined\ in \ this \ work},
 \eea
  where  ${\cal O} [\bar g_{ij}]$ stands for all possible functionals of $\bar g_{ij}$, expressed in any given set of coordinates.  The number of degrees of freedom of  $\bar g_{ij}$ modulo the internal reparametrization symmetry is $
 \frac  {d (d-3) }{2}$.  As  it should, this number corresponds  to  that of a physical graviton in dimension $d$.  The right hand side of~(\ref{?}) will be expressed by an  appropriate use of the BRST symmetry for the unimodular  gauge choice within the ADM  foliation framework (see the precise expression \eqref{ansatz}).  Therefore, the use of the ADM  decomposition  of the  pseudo Riemannian  $d$ metric field for $d>2$  clarifies enormously   the task of representing  the gravitational observables,  a bit  analogously   as the   use of  the Beltrami    parametrization of    Euclidean $2d$-metrics clarifies  the handling of $2$-dimensional  quantum gravity and of its conformal properties~\cite{ Baulieu:1986hw}.

Eventually, 
 the cohomology     selecting  the observables~(\ref{?})     is made of  the  classes of equivalences of    space time metrics    $\gmn$  represented   by  the leaf metrics $ \gij$  defined    modulo   Weyl and    leaf reparametrization  symmetry. Making  explicit its representants     is  made possible by   the separations  of  the  lapse and shift functions introduced in \cite{Arnowitt:1962hi} and the use of the unimodular gauge.  
This definition of observables is compatible with   {  the definition } of  the classical degrees of freedom  in     \cite{York}.  So, the gauge fixing exhibited in this work      conveniently  selects  the appropriate quantum field theory   field variables including the relevant ghosts. Moreover, the paper shows that the chosen  gauges  for the ADM  shift fields  share   analogies  with the       regularized  Coulomb gauge of the~Yang--Mills~theory \cite{lbdz}.
 

\section{Preliminary remarks}
 The  Einstein  action including a cosmological term and matter interactions  is 
$S=\int d^d x\sqrt g [R^d(g_{\mu\nu}) +\Lambda  +   { matter\ field \ interactions}]$ in a $d$-dimensional Lorentzian curved space-time   ${\cal M}_d$ {{} with metric $g_{\mu\nu}$ and  scalar curvature $R^d (g_{\mu\nu})$}.
The    proposition of York    \cite{York} is   that 
the physics of classical  gravity is carried by the conformal classes of the 
metrics of spacelike $d-1$  dimensional leafs $\Sig_{d-1}$ in
 of  ${\cal M}_d$, 
 a property  that is quite     transparent  by defining  \`a la ADM the possible foliations    of  ${\cal M}_d$  in a    {coordinate} reparametrization invariant way.

  The geometry  of  the  { pseudo Riemannian  } manifold   ${\cal M}_d$ is encoded  in the brute force  knowledge  of  the
$   \frac{d(d+1)}{2}$ component of  its metric tensor $g_{\mu\nu}( x^i,t)$ (minuscule latin indices such as $i$ denote  spatial coordinates and greek ones world coordinates, with $x^0\equiv t$). The   reparametrization invariance   allows one to use     any      given {     {  consistent   }} set of 
coordinates $\{  x^\mu   \}$.  However  it   also implies  a certain  analytical fuzziness that is  often  harmful to define physics. It makes  it     not obvious   to   formulate  the proper definition    of the physical observables at the quantum level (even assuming that  the   quantisation scheme itself were well defined).

 The   $ \frac{d(d+1)}{2}$     field  equations   of $g_{\mu\nu}$   
  build a    non linear second order in time  $t$   Cauchy problem. Given suitable initial conditions, only  
 $  \frac{d(d-1)}{2}$ independent combinations of the $ \frac{d(d+1)}{2}$  solutions truly propagate a physical dynamics.   The 
 remaining independent $d$ combinations are       constraining identities with no dynamics.
 This     property  can be   explained  in  the  Hamiltonian  framework:     only the~$  \frac{d(d-1)}{2}$~components $g_{ij} $
  of    each  foliated leaf    metric    have  classically  propagating Hamiltonian momenta  $\pi_{ij}$; 
 the $d$~other components $g_{\mu 0} $    have no Hamiltonian momentum. This coincides with the fact that the~$ \frac{d(d+1)}{2}$~solutions for $\gmn$  are   not independent. 
The difficulty of dealing with the constraints in the gravity  theory   surpasses the   analogous one   in   the Yang--Mills theory.  The later     is also a constrained system. The   temporal component $A_{0}$~(the formal analog of $g_{\mu 0}$) of the Yang--Mills  field has no momentum,  but   the   problems  that it causes can be solved by a rather  elementary gauge fixing  (including a simple recourse to the notion of BRST symmetry).         The   Yang--Mills  observables can be  then clearly defined  as  gauge covariant functionals of the gauge fields.    In the~gravity case, the constraints    mix with the dynamics, leading to the   time problem widely discussed in the literature.   { In fact, with no preferred  time, it is not an obvious task to define the notion of a  time ordered path integral independent of the  choice of  the  system of coordinates,  contrary to what happens    in the Yang--Mills situation.}    {  Many  of the   difficulties  that  physicists  encounter when facing the quantum gravity theory are partly due to the fact that  they   push  often  too far  the comparison between the notions of  the reparametrization symmetry and  that  of a gauge symmetry, that is, beyond the level of local Lorentz symmetry.}

 Solving  the complexity
  of the constraint  problem   in   theories with a   field reparametrization invariance  often  requires a more sophisticated determination  of  the  BRST invariance of the quantized theory   than in theories with a standard gauge invariance. 
  Topological quantum field theories, where one starts from a classical theory defined by a purely boundary term, and thus  with a vanishing Hamiltonian, are in fact   a good example of this sophistication.  New fields belonging to unphysical trivial quartets  of  the BRST symmetry  must be often introduced to solve the quantization of systems with imbricated   local symmetries. This causes a  non trivial  departure from the purely    { classical} concepts. These     unphysical fields    undergo themselves a non trivial propagation  to ensure a unitary behaviour of the  physical sector   in the context of a local quantum field theory.
 
 To summarize these remarks, the $d$ gravitational constraints stemming  from  the gravity action  are   reparametrization covariant.
  When they are expressed      { in terms} of the independent   $\frac  { d (d+1)}{2}$  field components $g_{\mu\nu}$, their expression is   rather  complicated, so   they  mix  non trivially with the equations for  the genuine propagation of physical degrees of freedom. The way it   occurs  strongly depends on the chosen gauge fixing.   This is basically what   makes   the   identification of   the  physical degrees of gravity quite   subtle.  In the Yang--Mills  case,   the  constraints amount to  the  quite simple Gauss law and     the  definition of the observables shows up  immediately   as   all possible correlators  of  the electric and magnetic fields (more generally as   all gauge covariant functionals of the gauge field). 
  The complexity of the   methods required in order to   define the  observables in the  Einstein theory contrasts with 
  the   relative simplicity of    those   that one can  use successfully
  in the Yang--Mills theory. 
 \section {ADM  parametrization  and some notations}

The ADM foliation representation~\cite{Arnowitt:1962hi}  of the $\frac{d(d+1)}{2}$ components of  the $d$ dimensional~metric $g_{\mu\nu}$ of  the Lorentzian space ${\cal M}_d$  is most appropriate    to  
compute    its foliation into  internal   spacelike $d-1$~dimensional leafs. Classically, 
the    Lorentz time is    
the  foliation parameter for   describing  the  dynamics led by the  $d$-dimensional Einstein
equations of motions. 
 This is  at the basis of the interesting notion of ``geometrodynamics"  of the   early  60's.   \cite{York} gave 
 a proof    that      the    classical observables of    gravity are in fact the conformal classes of  $d-1$~dimensional metrics of constant~$t$~leafs in ${\cal M}_d$. This is a remarkable result since      the  theory is  based on the non  Weyl invariant  Einstein
equations of motion.  
To unravel the possible  Weyl covariance  properties of the theory, and implement   unimodular decompositions,    the recent work of \cite{Kiefer:2017nmo} is  extremely useful    in the
spirit of \cite{Thomas199,  Thomas2,  Thomas3}.
 In~\cite{bcw} the unimodular decompositions  have been  shown in the    different context of stochastic quantization to be   implemented  by
a  ``golden rule'' that  makes easier   the       decomposition of   all tensors 
quantities as  function of Weyl invariant $\Sig$-tensors plus terms
depending on the conformal factor.

The ADM
parametrization \cite{Arnowitt:1962hi} consists to express as follows  the  squared   invariant  Lorentz length
\bea\label{ADM}
\d s^2=-N^2\d t^2+(\d x^i+N^i\d t)g_{ij} (\d x^j+N^j\d t)
\eea
of   any given infinitesimal line element in the    space ${\cal M}_d$ with coordinates $\{    x^\mu    \}   $. The $d$-dimensional metric~$^{d}\gmn$~with
  $ds^2={^{d}\gmn} \d x^\mu \d x^\nu$ 
   has Lorentz signature  $(-,+,\cdots,+)$.
  The lower and upper  indices components~$^{d}\gmn$ and  $^{d}g^{\mu\nu}$  of the metric are related  to the foliation framework parametrization  as follows 
   \bea\label{admf}
   ^{d} g_{\mu\nu} =
    \begin{pmatrix}
    g_{ij} &N_i
    \cr
    N_j &   -N^2+N_i N^{j}
     \end{pmatrix}
     \quad
     ^{d} g^{\mu\nu} =
     \begin{pmatrix}
    g^{ij } -\frac {N^i N^{j} }{N^2 }  &\frac {N^i }{N^2}
    \cr
   \frac {N^j }{N^2}
 &  - \frac {1 }{N^2}
     \end{pmatrix}.
  \eea
   $N^i(x^j,t)$  is 
  the  shift vector,  $N_i \equiv   g_{ij} N^j$ and
$N(x^j,t)$ is the lapse function.   $N^i$  is Weyl invariant because   $\d x^i+N^i\d t$ has Weyl weight zero. The  $d$-dimensional  metric  determinant  is  $g\equiv \det(^{d}\gmn)=-  N {\sqrt {\det {g_{ij}}}}  $,   
$\det (g_{ij})>0$.
\def\xmu{   \{  x^\mu\}  }
 \def\bm{  {\bf N}}
  \def\bmn{  {\bf n}}

In the  ADM  framework, one  considers $(g_{ij,}N,N^i)$ as fundamental fields. 
Each  $d-1$  dimensional leaf  $\Sig_t$ of the foliation is defined for  a fixed  given value of  
  $t=x^0$. 
$\Sig_t$~has internal metric $g_{ij}(x,t)$ with Euclidean signature.        One has  a  normal vector $
{\bm }\equiv N^{\mu}\pa_{\mu}=\frac{1}{N}\pa_0-\frac{N^i}{N}\pa_i $  at  each point $x^i$ of the leaf, pointing out 
in  the extrinsic  space  of $\Sig_t$ in  ${\cal M}_d$, and  a          
      one-form   $
\bmn  \equiv N_{\mu}\d x^{\mu}=-N\d{t}
$.
The normalisation   is  $N_{\mu}N^{\mu}=-1$.
  \cite{Gourgoulhon:2005ng, Gourgoulhon:blau}    define a  useful   projector $P^{\mu}{}_{\nu}$  on $\Sig$   that  extends  covariantly  to the
  space  ${\cal M}_d$ all  objects defined on  $\Sig_t$. It is defined as
\bea
P^{\mu}{}_{\nu}\equiv \delta^{\mu}{}_{\nu}+N^{\mu} N_{\nu},\quad
P^{\mu}{}_i g_{{\mu}j}=g_{ij},\quad
P^{\mu}{}_{\nu} N_{\mu}=0.
\eea
 
\subsection{Leaf covariant tensors}
The extrinsic curvature $K_{ij}\equiv\demi\Lie_{\bm}\,g_{ij}$ at each  point  of a leaf   $\Sig_t$
is
\bea\label{Extri}
K_{ij}=
\demi\big(N^{\rho}\pa_{\rho}g_{ij}+  g_{{\rho }i}  \pa_j   N^{\rho}+
 g_{{\rho }j}  \pa_i   N^{\rho}
 \big)
 =\frac{1}{2N}
\big(\pa_{0} g_{ij}-   
 N^k\pa_k g_{ij}
      -   g_{kj}\pa_i N^k
- g_{ki} \pa_j N^k \big)=
\frac{1}{2N}(\pa_{0} g_{ij}-\nabla_i g_{j{t}}
-\nabla_j g_{i{t}}).
\eea The dependance of the extrinsic curvature $K_{ij}$ on       $ \nabla_{N^k} g_{ij}\equiv {\Lie}_{N^k} g_{ij}=   N^k\pa_k g_{ij}
      +  g_{kj}\pa_i N^k
+ g_{ki} \pa_j N^k
      $       makes explicit   that it     is   a   covariant   entity in the leaf    
orthogonal   to $\bm$.
     Using  $P^\mu_\nu$,     $K_{ij}$ extends in ${\cal M}_d$ as
 \bea
K_{{\mu}{\nu}} \equiv P^i_\mu P^j{}_{\nu} K_{ij}\quad 
\rm {so} \quad  K_{{t}{t}}=N^i N^j K_{ij}& &
K_{{t}i}=N^\mu    K_{\mu i}=N^\mu  K_{i\mu }=K_{i{t}}.
\eea
The $N$-independent  covariant  speed 
$D_{0} g _{ij}$ of a leaf along its normal is
\bea\label{speed}
D_{0} g_{ij}\equiv 2NK_{ij}=
\pa_{0} g_{ij}-    
 N^k\pa_k g_{ij}
      -   g_{kj}\pa_i N^k
- g_{ki} \pa_j N^k .
\eea
It is    in fact    interesting  to  introduce the leaf covariant  rate of evolution   $  a _{ij} $  of the speed $K_{ij}$,  
 \bea\label{acce}
  a _{ij}\equiv N\Lie_{\bm{}}(D_{0} g _{ij})=(\pa_{0}-N^k\pa_k)D_{0} g_{ij}-2 D_{0} g_{k(i}\pa_{j)}N^k.
\eea
This  acceleration $  a _{ij} $  of   the leaf along its normal~$\bm{}$
 is the   part of the  $d$ dimensional 
Riemann tensor $R^\mu_{\nu\rho\sigma}$ that is covariant in the leaf at constant $t$
and contains the term $\pa_{0}^2g_{ij}$ but no derivative of the lapse
function $N$. This tensor is a  useful leaf covariant entity  to study the  various non relativistic approximation of general relativity.

\section {Defining the  classical and quantum evolution}

We will    define the gravitational physical degrees of freedom  of the gravity field as the unimodular part of $g_{ij}$, denoted as $\bar g_{ij}$ with  $\det \bar g_{ij}=1$.  One has the relation  $\bar g_{ij} =\frac { g_{ij} }  { (\det  g_{ij} )^{\frac 1{d-1}} }$.  When  $\bar g_{ij}$ is defined modulo internal reparametrization,  it counts  the right number of degrees of freedom of gravity quanta in each leaf,  that is,
$\frac {d(d-1)}{2}-1 -(d-1)=\frac{d(d-3)}{2}$.  
For the special case $d=2$, one rather expresses  the 2 dimensional metric  by its Beltrami differential and  conformal factor   and the relevant field of  $2d$ gravity is its Beltrami differential  \cite{Baulieu:1986hw}. 
Note that for  $d=3$ one has $ \frac {d(d-3)}{2}=0$ and $3d$ gravity is indeed topological. The conformal classes of its  $2d$ spatial leafs 
 are then represented by their moduli, whose finite  ranges can be restricted to fundamental domains.

 A first    question is     to   better understand the   classical $t$ evolution of the leafs $\Sig_t$ defined  modulo local dilatations through their $\bar g_{ij}$ dependance.   \cite{York}  did so      by using extensively    the foliation concept.  

A second  question  is  to    define    the    expectation values of functionals of  these physical degrees of freedom.  It  requires  a plausible definition of the amplitude of probability 
  \bea
 \label{calA} 
  {\cal {  A}} (\Sig_I\to \Sig_F) \quad  \rm {for} \quad  t_I  \to  t_F   \eea
 of going from a given leaf  $\Sig_I$ at time $t_I$ to another given leaf $\Sig_F$  at time $t_F$.
 The  definition of $ {\cal {  A}} (\Sig_I\to \Sig_F) $ will follow  
 from a  suitably gauge fixed  BRST invariant    path integral formula that we will introduce for       observables. 

    A third question is   how to  realistically   compute such     averages  of   functionals of the $\bar g_{ij}$'s  once they have been defined as the observables.  But then, one    expects  to hit all known local quantum field theory  difficulties.    In particular, one has  those coming from the sign changes  of     the $d$~dimensional Euclidean Einstein action    when performing     the   summation  over all  metrics in the path integral. Then stochastic quantization might be useful~\cite{bcw}.
    

 What is wonderful for both first and second questions  is that the use of the ADM parametrization of the metric tensor $g_{\mu\nu}$ changes the conceptual  visualisation of the space $\xmu$,     as an ordered accumulations  of leafs rather than   a  disorganised ensemble of points. From a physical point of view, 
 the foliation  picture  is such that, within  the same system of coordinates,  one can  heuristically  change the foliation by  local  small dilatations of neighbouring leafs   along their normal and by local small shifts without changing   the physics, while maintaining the
 $d-1$~reparametrization invariance  $\Diff$ in each leaf. 
 
 This property is often referred to by the sentence, ``the choice of slicing of space-time does not matter".  One of its illustrations  is that one can solve the Cauchy problem of the gravity theory by adjusting at will the lapse and shift functions  in each infinitesimal step. The role of a sophisticated enough BRST symmetry is to impose  this  property directly at the quantum level, while allowing the definition of a consistent local gauge fixed action.

\subsection{Classical  action and the positivity  property of its ``kinetic energy" term}
Both the speed $D_0 g _{ij}$ and the acceleration ${\A}_{ij}$ are covariant tensors in each leaf with respect to $d-1$~dimensional diffeomorphisms with $t$-independent parameters $\xi^i(x)$, denoted  as $\Diff$.  $N$ and $N^i$ are respectively a scalar and a vector for such diffeomorphisms.
Using the ADM field variables one has for the cosmological term  $
\sqrt  { |g|} \equiv \sqrt {-\det g_{\mu\nu}    } =N  \sqrt {\det g_{ij }} $
and for the Einstein-Hilbert  action 
\def\dW{\lambda}
\bea\label{admE}
 \int \d ^d x  R^{d} ( g _{\mu\nu} )=
\int 
 \sqrt {       \det g_{ij}        }\ 
     \d^{d-1}x \ \d t  
 \Big (
    \frac {g^{ijkl}   }{N} \ D_0 g _{ij}          D_0 g _{kl} +   {N}  R^{d-1} ( g _{ij} )
 \Big ) 
 \eea
  Here, $g^{ijkl}  =    \frac{1}{2} (g^{ik} g^{jl}+ g^{jk} g^{il})  -\dW g^{ij} g^{kl} 
    $  is the (so-called DeWitt)  leaf metric  4-tensor  over the space of  Euclidean  $d-1$~dimensional metrics. 
    The constant $\dW$    is to  be  computed  by the decomposition $ R^d=    \frac {1}{N^2} \ D_0 g _{ij}    g^{ijkl}         D_0 g _{kl} +   R^{d-1} ( g _{ij} )$.
    $g^{ijkl}$ has $\frac{d(d-1)}{2}$ diagonal  elements  with             signature      $(1-\dW(d-1),1,\ldots,1)$.
   This signature is    non  Euclidean for all values of $d>2$ (although  
   $g_{ij}$ is Euclidean). Indeed the computation shows that     $1-\dW(d-1)<0$  generically.  
   The   non positivity of the kinetic term  can be seen  by  decomposing  
   $g^{ijkl}$ in traceless and trace parts
   \bea g^{ijkl}  = (g^{ijkl})^T   +\frac {1-\dW(d-1)}{d-1} g^{ij} g^{kl}. 
 \eea
 Here the   $\lambda$ independent   traceless part of $g^{ijkl}$ is 
   \bea
   (g^{ijkl})^T= 
    \frac{1}{2} (g^{ik} g^{jl}+ g^{jk} g^{il})  -\frac {1}{d-1} g^{ij} g^{kl}. 
 \eea
        Besides, $  K^T_{ij} =K_{ij}-\frac{1}{d-1} g_{ij} K
      $ is traceless  where       $K\equiv g^{ij } K_{ij}$.
      One has  thus
      \bea\label{dW}
    g^{ijkl}       K_{ij}K_{kl} 
      &=&   K^T_{ij}   K^{T{ij}} + \frac {1-\dW(d-1)}{d-1} K^2\quad{ \rm where}\  K^{T{ij}} \equiv g^{ik} g^{jl}K^T_{kl } .
      \eea
    With  $ K _{ij} =\frac {1}  {2N }D_0 g _{ij}  $ and   $1- \dW(d-1)<0$,      Eq.~(\ref{dW}) implies  that 
        the traceless and    trace parts  of the ``kinetic energy term " 
      $   g^{ijkl}   D_0 g _{ij}           D_0 g _{kl}$  
       have opposite signs. 
       One  motivation for  
          the unimodular  gauge  choice  we will shortly   impose in a BRST invariant way is  then quite clear.     The gauge choice  
          $\sqrt {\det   g_{ij}}=1$  implies    $g_{ij} =\bar g_{ij}$ and   $K|_{g_{ij}=\bar g_{ij}  }=N g^{ij}D_0 (\bar g_{ij})=0 $ since   the trace  of any given variation of an unimodular metric  vanishes. So, 
            \bea \label{magics} \frac { g^{ijkl}  }{N} D_0   g _{ij}        \  D_0   g _{kl}  |_{  g _{ij} =\bar g _{ij} }
   =  \frac {1}{N} D_0 \bar g _{ij}  \      D_0 \bar   g ^{ij}  >0
     \quad{ \rm where}\  D_0 \bar g ^{{ij}} \equiv \bar g^{ik} \bar g^{jl}D_0 \bar g_{kl }.
   \eea    
     The  "kinetic energy term"  
    $\frac {1}{N} D_0 g _{ij}    g^{ijkl}         D_0 g _{kl}$  is   thus  a sum of 
     positive terms   in each       leaf  expressed in the unimodular gauge. 
      It must be noted that the negative  direction in the space of deformations of spatial metrics  is      associated with overall volume deformations. The  choice of the  unimodular gauge   is     quite suggestive from  this point of view, as   it implies that all deformations of the metrics be   purely traceless.
As it is well known    the lapse and shift $N$ and $N^i$  have no genuine dynamics since (\ref{admE})  has  no dependance in their time derivatives, so this picture remains true for all possible choices of   $N$ and $N^i$.
\footnote{
\bluecor
The positivity of the  ADM kinetic energy   the unimodular gauge is an important  point, but  more is needed to prove  that the standard conformal mode instability in Euclidean quantum gravity can be  cured. Indeed,  although
the elimination of   the  modes that  comes with the wrong sign
is granted in the unimodular gauge, this does not resolve entirely the problem since the gauge-invariant scalar mode in a York-like decomposition still has the wrong sign \cite{York}.   A  careful analysis  of possible  compensations from the  ghost  dependence of BRST invariant action is needed   to help solving this question.
 The    relation between  this question and the BRST framework developed here  will be the subject   of another publication. }

\subsection{Using the leaf covariant tensors in  classical gravity}

 $D_0 g _{ij}$  defines  the  classical   evolution of the leaves. Both $D_0 g _{ij}$  and ${\A}_{ij}$ are  independent on $N$.  This sustains   the     importance  of the    foliation language in gravity,  for a   step by step   evolution of leafs,  with an arbitrarily given choice of   $N$ and $N^i$ at each step, while remaining in the same system of coordinates $x^\mu$.  Suggestively, a  local change  of the lapse can be compensated   by a local dilation of $g_{ij}$, so a local change of $N$ is of no relevance for the conformal   classes of  leaf internal $d-1 $ dimensional metrics $g_{ij}$. Changing the shift vector is also classically  irrelevant for the evolution of these  conformal   classes.
  In fact, the independence  of the physical evolution   of the gravity field under changes of the initial  lapse and shift functions that  is explicit  in the work of York  has been   widely used in the context of numerical gravity.  It allows   to possibly    optimize  the stability of   numerical computations  since a blind  simulation implies  more systematic errors in the enforcement of  gravitational constraints.

 \def\cL{  {\Lie}}
 \def  \Md {  {\cal M}_d} 
 \subsection{Using the leaf covariant tensors for a proper definition of the path integral}
   
Classically, the  boundaries of the Cauchy problem of   Einstein equations can be chosen as     the     metric $g^I_{ij}(x^k, t_I)$ and       extrinsic curvature  $K^I_{ij}(x^k, t_I)$ (basically the initial  speed of the leaf)
 of    an initial spacelike  leaf   $\Sig_I$. 
  The classical limit suggests      using     $\phi= \log  \sqrt {\det g_{ij}(x^k, t) } $ and    the  $\frac { d(d-1)}{2} -1 $ independent components of the  unimodular    $ \bar g_{ij}(x^k, t)   $    as   the     leaf   field variables  rather than the      $\frac {d(d-1)}{2} $   components of  $g_{ij}$'s.   
In each  leaf of the evolution, one has the internal  diffeomorphism symmetry  $\Diff$  
$x^i\to x'^i(x)$, whose infinitesimal transformations are represented
by the Lie derivative $\Lie^{\Sigma}_\xi$ along a ${t}$-independent vector field
$\xi^i(x^j)$. 
 The  differential operator $\Lie^\Sigma_\xi$ acting  on leaf tensors only involves the partial derivatives $\pa_i$'s and the $g_{ij}$'s.
 $N^{i}=g^{i{t}}$ and  $N   $  transform   as  a  $d-1$ vector and a  $d-1$  scalar   
 under $\Diff$. The transformation laws  of  $g_{ij},N^i,N$    under the complete    reparametrization symmetry   in     ${\cal M} _d$        is defined  from  $s^{d}\gmn   =    {  \cL _{\xi^{^\rho}}  }  ^{d}\gmn$
 and the chain rule applied to  the relations~(\ref{admf}). Their parameters and ghosts are     $\xi^i(x^j) \to    \xi^\mu (x^j,t) = (\xi^0 (x^j,t)   , \xi^i(x^j,t)) $.
    The action of $\Lie^\Sigma_\xi$ is thus completed  in ${\cal M} _d$ by terms    involving   $\xi^t\equiv \xi^0$ and transformations  under $\pa_0$.   Explicit formula will be shortly~displayed.
 
 We can now    tentatively  define the path integral that computes the amplitude of probability ${\cal {  A}} (\Sig_I\to \Sig_F)$~introduced in~(\ref
  {calA})
 of going from a given leaf  $\Sig_I$ at time $t_I$ to another given leaf $\Sig_F$  at time $t_F$.  The boundaries are thus defined as both   $d-1$ dimensional metrics of initial and final leafs.
 The guidelines are as follows.

 At  the quantum  level, $ {\cal {  A}} (\Sig_I\to \Sig_F) $ is non zero   when  the path in the metric space $\{g_{ij} \}$ that connects $\Sig_I$~to~$\Sig_F$ is not a solution of the equation of motion, although it should  concentrate around it  when    $\hbar\to 0$.
 
  The leaf evolution can be  decomposed  into  N steps,  each  one  becoming an infinitesimal one in  the limit $N\to \infty$,   with  the   following composition   formula, to be  read from right to left 
 \bea\label{path}
  {\cal {  A}} (\Sig_{I}\to \Sig_{F})
=
 \int \int  \ldots\  \int  [\d  g_{ij}^{(n-1)}]   [\d  g_{ij}^{(n-2)}]      \ldots[\d  g_{ij}^{(1)}] 
 {\cal {  A}} (\Sig_{n-1}\to \Sig_F)
  {\cal {  A}} (\Sig_{n-2}\to \Sig_{n-1})
 \dots
  {\cal {  A}} (\Sig_{I}\to \Sig_{1}).
   \eea
   This formula must remain consistently true  for $N\to \infty$, meaning that each  intermediary leafs become infinitesimally  near   its  right and left neighbours.
   
   The goal is to provide    a consistent   definition of  the gravitational observables. One expects
  \bea\label{try}
 <{\cal O}  (\bar g_{ij})  > \equiv \int  [\d  g_{kl}]  [\d N]  [\d N^k] 
 [\d \rm( {ghost]   [\d \rm( {antighost }]    [\d (Lagrange \ multiplier})]
 \nn\\
  {\cal O}  (\bar g_{ij}) \exp \int -\frac {i}{\hbar} S^{BRST} [\bar g_{kl}, \rm {ghost, antighost, Lagrange \ multiplier}].
 \eea
  In this definition,  $S^{BRST}$ corresponds to  a   BRST invariant action with a   gauge fixing  of the  zero modes of the Einstein action, obtained by adding a BRST exact term to the classical gravity action.
    We will shortly define what are all ghosts and antighosts  in(\ref{try}).  In  the foliation picture, 
    the  BRST symmetry  in (\ref{try})     involves at least  
 the   $d$-vector ghost $\xi^\mu  (x^j,t) \equiv (\xi^0,\xi^i)$.   
   Other ghosts are needed in  the unimodular gauge.  $S^{BRST}$  must be such that all observable mean  values  $ <{\cal O}  (\bar g_{ij})  > $  are     covariant under  the full $d$-space coordinates transformations $x^\mu \to x^\mu + \epsilon ^\mu$. Once  $(\ref{try})$   will be made precise, and  if proven consistent, we expect  it will define
 $ {\cal {  A}} (\Sig_{I}\to \Sig_{F}) $.

   \subsection {BRST approach}
   
       The equations of motion   of the   Einstein action in the ADM variables  $ \int 
 \sqrt {      \det g_{ij }        } 
  \d^{d-1} x \d t  
   (
 \frac 1 N    g^{ijkl}  D_0 g _{ij}            D_0 g _{kl}  +   {N}  R^{d-1} ( g _{ij} )
    )$
    show     that there is no classical dynamics for the lapse $N$  and shift $N^i$ at the classical  level.  
   The path integral  must  sum over      all possible     evolutions of   leafs  identified  by   the  field variable  $g_{ij}$ modulo a gauge fixing for the  reparametrization  symmetry in the leaf, between   $g_{ij}^I$
     at  initial time $t=t_I$ 
  and   $g_{ij}^F$ at  final time $t=t_F$. 
 Among all path connecting $g^I_{ij}$ and $g^F_{ij}$,    only a single one    satisfies the Einstein equation, with   a determined   value of  the extrinsic tensor $K^I_{ij}(x,t_I)$   at the initial time.

The expected consistency is from  the  reasonable assumption   that the configuration space  of $d-1$  dimensional metrics  over which one      performs a path integration between $g^I_{ij}$ and $g^F_{ij}$  is a complete one for each value of $t$.  Indeed  the set of all metrics $g_{ij} (x^k)$ defined   modulo  $d-1$  dimensional reparametrization symmetry        describes   all possible $d-1$  dimensional manifolds $\Sig$ that may occur at   any given intermediate  value of the time $t$.  It  is in fact our freedom to  postulate   the definition of the path integral over all possible intermediate 
 leafs that may occur for the quantum evolution between $\Sig_I$~and~$\Sig_F$. 
 But, once it is stated,  the  consistency   of the definition  must be checked. In particular,
  one must ensure the    compatibility  of the path integral  definition with everything      we know to  be  true       in the classical limit as expressed   in \cite{York}. Since one  sums over all  intermediary  $d-1$ dimensional metrics in each of the     steps $1,2,\dots, n-2,n-1$, one must    ensure  that one can reinitialize {{the lapse}} at each step, meaning for instance that one   can impose
 \bea
 N_{n-1} =N_{n-2}\ldots =N_{1}= N_{I}.
 \eea
 This requirement  will be proven to be possible by a BRST  invariant gauge-fixing procedure.  Appendix \ref{App1} shows that 
 the BRST invariant gauge fixing we will do for  the shift field $N^i$   is analogous to  what one  does  for  gauge fixing  the temporal component of a gauge field  with a regularized Coulomb  gauge condition $A_0 =\A \pa_i A_i$. 

 Let us begin by rewriting   a bit more precisely the path integral proposed formula  (\ref{try}):
  \bea\label{ansatz}
 <{\cal O}  (\bar g_{ij})  > =\int  [\d g_{kl }] [\d N]   [\d N^k [\d  \rm ({ghost})]    [\d  \rm ({antighost})] 
  [\d  (Lagrange \ multiplier)]) 
  \ 
   {\cal O}  (\bar g_{ij})
  \nn\\ \exp  \int -\frac {i}{\hbar} 
         \int^{t_F, g_{kl}^F}_{t_I, g_{kl}^I} \d^{d-1} x \d t 
 \sqrt {      \det g_{kl }        }\ 
    \Big [
  \frac {1} N g^{klmn}   D_0 g _{kl}           D_0 g _{mn }  +   {N}  R^{d-1} ( g _{kl} )\nn\\
  +   {\blue {{\rm{BRST\ exact\ term} \ (ghost,  antighost,  Lagrange \ multiplier}, g_{kl}, N, N^k     )}}
  \Big ]
 \eea
  The Einstein action   in the second line   is invariant under  the full  ${\cal M}_d$  reparametrization symmetry  
       $s g_{\mu\nu}   ={  {\Lie}} _{\xi^\rho} g_{\mu\nu}
       $, 
         with  $g_{\mu\nu}= (\bar g_{ij}, {\sqrt {\det {g_{ij}}}}  ,N,N^i  _I  )$. 
Therefore, the ghost sector includes     at minima   $d$-space reparametrization  anticommuting   ghosts and antighosts  $d$-vector
 $\xi^\mu=  (\xi^i, \xi^0 )$ 
 and
   $\bar\xi^\mu=  (\bar \xi^i,\bar  \xi^0)  $ 
   with Lagrange multipliers
    $b^\mu=  (b^i, b^0 ) $,  
        $  s\xi^\mu=\xi^\nu\pa_\nu\xi^\mu$,  $s\bar\xi^\mu=b^\mu$~and~$sb^\mu=0$. The  $i$ and $0$ components of those fields  are respectively    scalar and  vectors with respect to $\Diff$.
  Other ghosts and  Lagrange multipliers      will be introduced as trivial BRST quartets as in \cite{unimodular} to enforce the unimodular gauge both in   $\Sig_t $ and in  ${\cal M }_d$, because  the transverse and longitudinal parts of the vector ghosts and antighosts must be distinguished in this case.  
 
The    BRST  symmetry operation $s$  is  thus the  nilpotent differential operator that  one can associate to   the 
 $\Md$~reparametrization symmetry.  Finding a suitable gauge fixing  must  be adapted to the   foliation scheme, meaning  choosing relevant gauge functions possibly allowed by the addition of  a   new   trivial  BRST field quartet. The action of $s$ on the leaf variables is obtained by using the relations (\ref{admf}). 
  An  equivariance  of $s$   with respect    to each  leaf  internal  symmetry    $\Diff  $ must appear. It is    explicit  by the leaf identification   of   $g_{ij } $ as an  internal   leaf $d-1$ dimensional  metric,  of      $N^i $  and  of  $\xi^i$  as   leaf  $d-1$   vectors,     
 and $N$ and $\xi^0$  as leaf  scalars.

 The unimodular leaf gauge  choice   is classically well defined   with the   following   gauge functions    
 \bea \label{unimodulargauge}
 N=1 \quad
\sqrt {- g} =   1 \quad        \pa^i   \bar g_{ij}  =0.
 \eea
 This generalizes          \cite{unimodular} within    the ADM parametrization.    Imposing  $N=1$ and $\sqrt{-g}=1$ is  as  imposing    $N=1$ and ${ {\det {g_{ij}}}}=1$ since $g =-N^2   { {\det {g_{ij}}}}$, and thus $g_{ij}=\bar g_{ij} $.   We will   shortly improve (\ref{unimodulargauge}) and  discuss  the gauge-fixing 
 of the shift $N^i$. 
 
 One could  refine the   gauge fixing of the lapse function,
$
 N=1\to N=f(x^i,t)
$, 
 where the background function $ f$ is   chosen at will, with $sf=0$, for instance $f(x)=N(x^i,t=t_I)$. Because  the   $f $ dependence would  be through a BRST-exact term, there should be no 
 $f $  dependence for the expectation values of physical  observables. For simplicity we will  stick to the choice $f(x^i,t)=1$.

 \subsection  {  BRST   spectrum for the  ADM parametrization unimodular gauge  fixing}

 The  unimodularity  gauge condition for the  metrics implies that one makes a  distinction between the longitudinal and transverse parts of the reparametrization  vector ghosts and antighosts   $\xi$ and $\bar \xi^i$ in the leaf.  Having  such a~decomposition  of vector ghosts modulo  longitudinal  parts  in a BRST invariant  way implies the necessity      
      of        ghosts of ghosts  \cite{unimodular}. The  longitudinal and transverse  components  of  the  Lagrange multiplier auxilary field $b ^i$  
must be also separated.    Altogether,  one needs    
introducing a  trivial BRST quartet  as in \cite{unimodular} \bea   \label{gg}    \L,  \ \eta  ^ {(10)},\bar \eta  ^ {(01)},\ b  ^ {(11)},
 \eea
which     completes  the ordinary  BRST system  for $\gmn,\xi^\mu, \bar \xi^\mu, b^\mu$. 
Having such a quartet   actually  
corrects the wrong statement  that       the  
 d+1 conditions  $\pa_i \bar g^{ij}=0,\ g=1$ and 
  $N=1$   might  imply    an over-gauge fixing.
The gauge function  $  \pa^i   \bar  g_{ij}  =0$   in (\ref{unimodulargauge}) will      be    smeared by  imposing  instead 
 $\bar g_{ij} \pa_k   \bar  g^{kj}  =-\pa_i L -{  {\alpha N_i}}$ and summing over possible configurations of  $L$, in the spirit of \cite{unimodular}.   $\A$ is some real parameter.
 In analogy with the case of the Coulomb gauge in the Yang--Mills theory (Appendix \ref{App1}), having $\alpha =0$ is a  singular gauge choice beyond the classical level. 
 One  thus generalizes   the leaf reparametrization gauge function in  (\ref{unimodulargauge})    by 
 \bea\label{unimodulargauge1}
    \bar g _{ij}\pa_k    { \bar g}^{k j}  =0  \quad\to \quad       \bar g _{ij}\pa_k    { \bar g}^{k j}  +\pa_i L+{  {\alpha N_i}} =0.
 \eea
   The    scalar  bosonic  fields  $L,b$ and     the  fermionic   fields  $\eta,\bar \eta$  will  count altogether for zero=1+1-1-1  degrees of freedom in the BRST unitary relations provided  their dynamics is governed  by  an s-exact  action with invertible propagators.
    Having  available  the  extra   unphysical fields   (\ref{gg})     turns out to be    what one  needs to get   a Lagrangian with    invertible propagators   and no zero modes  in   a BRST invariant gauge fixing for the    unimodular gauge.
  
       {\blue {  The following diagram displays suggestively all    necessary ghosts, antighosts and Lagrange multipliers\footnote{ The notation   $\phi^{g,g'}$  means that the field $\phi^{g,g'}$ carries ghost number  $g$ and antighost number $g'$ for a total  net ghost number  $G=g-g'$.  $\phi^{g,g'}$ is a boson if    $G$ is even   and a fermion if $G$ is odd. We often skip these ghost and antighost indices in the formula.}
        \def\red{\color{red}}
\bea\label{spg}  \quad \quad
\begin{matrix}
     &     \bar g_{ij},  {\sqrt {\det {g_{ij}}}} , N,   \L , N^i &   & &   \\
 \swarrow  \  \ \ \  \swarrow \  \ \ \  \swarrow& &    &  &   & & \\
    \xi^{i(10)},  \xi^{0(10)} , \eta  ^ {(10)}  &    &    \bar  \xi_i  ^ {(01)}, \bar  \xi_i0 ^ {(01)} , \bar  \eta  ^ {(01)} & \\
       &   \swarrow \ \ \  \swarrow \  \ \ \  \swarrow &  &    \\ 
     &     b ^{i (11)},  b^{0   {(11)}},  b  ^ {(11)}  &    &    \\ 
     &   &  &    & &   \\
  1   &   0 &     -1  & 
\end{matrix}.  \eea}}
The  numbers  $-1,0,1$   in  the  bottom line    indicate  the   net ghost number of      fields that are   aligned  vertically   above  each number.
{\blue {The     BRST transformations    are \def\Lx{\Lie_\xi}
 \bea      \label{aux} 
  sg_{\mu\nu}&=& \Lx   g_{\mu\nu}
  = \xi^\rho  \pa_\rho \gmn + \bar g_{\rho\mu}\pa_\nu \xi^\rho + \bar g_{\rho\nu}\pa_\mu \xi^\rho
   \nn\\
   s\xi^\mu&=&   \Lx   \xi^\mu =\xi^\nu\pa_\nu \xi^\mu
   \nn \\  
\s\bar \xi    ^\mu   &=& b     ^\mu    \quad  \quad \quad\quad \quad\quad \quad
    \s b  ^\mu   =0
     \nn\\
   \s L   &=& \eta      \quad \quad \quad \quad \quad\quad \quad
   \  \s \eta      =0
     \nn\\
      \s\bar\eta      &=& b    \quad \quad \quad\quad \quad\quad \quad
  \ \   \s b   =  0  .
      \eea
 The   operations $d$, $s$,  $i_\xi$, $\Lie_\xi=[i_\xi,d]$
 build    a system of nilpotent graded differential  operators acting  with $[s,\partial _\mu]\equiv 0$, $\{ s,\d\} \equiv 0$ and     
$  { s^2  =0 }$ on all  fields\footnote{ The convenient graded operation 
$\hat s=s-L_\xi$  is nilpotent in the absence of local supersymmetry because in this case  $  s\xi=   \Lx \  \xi$.
In supergravity, one  has also  $s^2=0$ with a complete system of auxiliary fields but then  ${\hat s}^2= i _\Phi\neq 0$ where $\Phi^\mu  =\chi\gamma^\mu\chi $ is  the vector field quadratic in the commuting  supersymmetry ghost $\chi$ \cite{bb}.}.
Both last   lines in Eqs.~(\ref{aux})  identify    $L,\eta,\bar\eta,b$ as the elements  of    a BRST exact quartet. The   extra commuting scalar $b $ in (\ref  {gg}) 
is   a    scalar  Lagrange multiplier with ghost number~0 and  both  anticommuting   scalars    $\eta$,   $\bar\eta$    
are    odd Lagrange multipliers with   ghost numbers $-1$ and $1$.}}

 One has $s\sqrt{-g} =\pa_\mu(\sqrt{-g} \xi^\mu)$ and   
the    $s$  transformation  of the unimodular    $\hgmn=  { \gmn}{    (\det \gmn) ^{-\frac 1 d} }$ is  traceless, 
\bea
 s \hgmn
=
  \hat g_{\rho\mu}\pa_\nu \xi^\rho + \hat g_{\rho\nu}\pa_\mu \xi^\rho -\frac {2}{d}  \hgmn  \pa_\rho\xi^\rho.
  \eea
{\blue {  The nilpotent   BRST $s$ transformations   of  $g^{ij}$,  $g^{0i} $ and  $g^{00}$  determine those of $N$,  $N^i$  and $\dg$ according~to  
   \bea\label{sN}   
s N
&=&
{\Lie}_{\xi^k}  N + \xi^0 \pa_0 N    + N  (\pa_0  -N^i\pa_i )  \xi^0
 \nn\\
   sN_i&=& {\Lie}_{\xi^k}  N_i +   g_{ij}\pa_0  \xi^j +\xi^0\pa_0 N_i +    ( N_i \pa_0 +(N^kN_k-N^2)\pa_i )\xi^0
   \nn\\
s \dg  &=&
\pa_k(\dg\xi^k)
+
 (\frac {1}{N}    -\dg )\pa_0 \xi^0+\dg (N^k\pa_k \xi^0 -\frac {1}{N} \xi^0\pa_0 N)
\eea
   where  ${\Lie}_{\xi^k}  N=  \xi^k\pa_k N$ and   ${\Lie}_{\xi^k}  N_i =  \xi^k  \pa_k N_i +N_k\pa_i \xi^k$.
    $s\bgij$ is      traceless and  lapse independent,   with  
 \bea
s  \bgij
= 
(
\xi^k\pa_k +\xi^0\pa_0      ) \bgij +
 \bar g_{ik}\pa_j\xi^k +\bar  g_{jk}        \pa_i\xi ^k
+   N_i\pa_j \xi^0   +    N_j\pa_i  \xi^0 
-\frac {2}{d-1} \bgij  (  \pa_k\xi^k 
+     N^k\pa_k \xi^0).
\eea }}
 \subsection  {The BRST invariant action in  the leaf unimodular gauge       }
 We can now express the $s$-exact  BRST invariant gauge fixed  action in  the refined leaf   unimodular gauge:   
   $$
   I_{BRST}
    ^{
    Unimodular
    }=
    \int\d ^dx     \sqrt {      \det g_{ij }   }     \Big[  
 \frac {1  } N    g^{klmn}    D_0 g _{ij}        D_0 g _{kl}  +   {N}  R^{d-1} ( g _{ij} )
  +{\blue {
 \ s    \Big( \  \bar \xi^0  (N-1)+    \bar \xi^i ( \bar g _{ij}\pa_k    { \bar g}^{k j}  {   {  +\A N_i}}+   \pa_i  L ) 
 +\bar \eta ( \sqrt {-g}-1) \Big)}}
\Big]$$
$$
=
  \int   \d t\ \d ^{d-1}x     
  \Big [ 
   \sqrt {      \det g_{ij }        }\ 
  \Big (
 \frac { 1  } N   g^{klmn}     D_0 g _{ij}   D_0 g _{kl}  +   {N}  R^{d-1} ( g _{ij} )
  \Big )
+   \    b^0 (N-1)  +b ( \sqrt {-g}-1) + b^i ( \bar g _{ij}\pa_k    { \bar g}^{k j}      {   { +\A N_i}}   +     \pa_i  L )
  $$
\bea
 -
  \bar \xi^0  s N  -   \bar \xi^i (s( \bar g _{ij}\pa_k    { \hg}^{k j}  {   { +\A s  N_i}} ) +   \pa_i  \eta ) 
 -\bar \eta \pa_\mu   \sqrt  {-g} \xi^\mu) 
\Big)
\Big].
 \eea
 Eliminating    $    b^0$ and $b$  by their  algebraic equations of motion imposes   $\sqrt {-g}=1,N=1 
 \to \gij=\bgij $.
  Thus
   {\blue {\bea\label{Egfl}
   I_{BRST}^{Unimodular}
   \sim
  \int   \d t\ \d ^{d-1}x     
  \Big[  
  \ 
  D_0 \bar g _{ij}          D_0  \bar   g ^{ij} + R^{d-1} ( \bar g _{ij}  )
  + b^i (  \bar g _{ij}\pa_k    { \bar g}^{k j} +   \A N_i  +
  \pa_i  L ) \nn\\
 -
  \bar \xi^0  s N  -   \bar \xi^i (s( \bar g _{ij}\pa_k    { \bar g}^{k j}    {   { +\A N_i}})+    \pa_i  \eta ) 
 +\bar \eta \pa_\mu    \xi^\mu) 
\Big].
 \eea}}
    Eq.~(\ref{magics})     has reduced  
        $D_0   g _{ij}    g^{ijkl}      D_0      g _{kl}|_{g_{mn} =\bar  g_{mn}}$ to 
         the positive   kinetic term  $ D_0 \bar g _{ij}          D_0  \bar   g ^{ij} =  g^{ik}g^{jl}    D_0 \bar g _{ij}        D_0  \bar   g _{kl} >0$ as  a   benefit  of using   the unimodular gauge.  
  One must    check all  propagators  are invertible.

 Consider    the bosonic sector. 
 The  $d-1$ components of the  Lagrange multiplier      $b^i$  gauge fix  the    transversality of the  quadratic approximation  of   $     R  ^{d-1} (\bgij) $ in function of $\bar g_{ij}$.
 This gauge fixing    involves 
      mixings with  $L$  and $N^i$ as seen in~(\ref{Egfl}). 
     The term $\pa_iL$ is justified   because the  eigenvalues      $\bgij $ depend  on only   $d-2$ independent combinations of the $\bgij $'s  due to   $\det \bgij =1$. The  $b^i$ propagation is  thus   between  
the  $d-2$~longitudinal degrees of freedom in   $\bar  g_{ij}$ plus    the  additional   one in $\pa_i L$   and  the    $d-1$  degrees of freedom  in $b^i$.

For $\A=0$, the propagation  of the shift  vector  $N^i$ is left undetermined, in analogy with the  time dependence of the temporal component of a gauge field that is left undetermined in the Coulomb gauge.

For  $\A\neq 0$, the elimination of $b^i$ gives a delta function 
 $ \delta(\A  N_i +   \bar g _{ij}\pa_k    { \bar g}^{k j}  +\pa_i L)$ in the measure   whose argument is linear in  $N^i$. One can then integrate over all possibilities over the lapse $N^i$,  meaning  that $N^i$ is gauge fixed at the zero of the delta function.
    The  gauge fixed  dependence in $N^i$  is   through  the  extrinsic  curvature   $    K_{ij}(\bar   g_{kl}, N^k)$
 gauge-fixed at $N=1$,       namely  through 
  $ {   K_{ij}}|_{N=1}=    D_0 \bar g _{ij} =  \pa_0 \bar g _{ij}
   - N^{(k)} \pa_k       \bar g _{ij}
    -  \bar  g_{ki}  \pa_j   N^{(k)}  
      -  \bar  g_{kj}  \pa_i  N^{(k)} 
      $ where $N_i=   -\frac{1}{\A} (   \bar g _{ij}\pa_k    { \bar g}^{k j}     +\pa_i L  )$. One gets     a propagator for the longitudinal part of $\bgij$ with a  quadratic behaviour in the time derivative of $\bgij$ and a quartic behaviour in  its  space derivatives.
  This propagator  is  analogous to  that of   $A_i$   in  the  Yang--Mills  regularized Coulomb gauge choice 
  $\A A_0 +\pa_i A_i=0$,   
  Appendix~\ref{App1} \footnote {  See \cite{lbdz} for   a detailed study of  the regularized Coulomb gauge  in the Yang-Mills theory.}
. 
       Thus, $\A\neq 0$  provides an   unambiguous unimodular    gauge fixing of each leaf. 

%

   Consider now the fermionic sector and    check    that the system of  the ghost propagators is also invertible. 
 It  is sufficient to investigate    the field quadratic approximation of the second  line in Eq.~(\ref{Egfl}).  
  In this approximation,     $
 \bar \xi^0 sN \sim\bar \xi^0  \pa_0 \xi^0 
$,  $  \bar \xi^i \pa _i sL+\bar \eta   s \sqrt {-g}
  \sim
     \eta   \pa_i \bar   \xi^i   + \bar \eta    \pa_\mu  \xi^\mu $ and
$ \bar \xi^ i s     ( \bar  g _{ij}\pa_k    { \bar g}^{k j}      +\A N_i )\sim
     \bar \xi^ i [ (   \bgij   \bar \Delta+ \frac{d-3}{d-1}  \pa_i\pa_j        ) \xi^j
  +\A   ( \pa_i \xi^0     +  \bgij       \pa_0    \xi^j)] 
   $.
   Thus the  quadratic approximation of the ghost action is 
    \bea 
 - \int   \d t\ \d ^{d-1}x \Big[  
  \bar \xi^0  s N 
   +  \bar \xi^i   \big (s( \bar g _{ij}\pa_k    { \bar g}^{k j}     {   { +\A N_i}} +     \pa_i  L)  \big )
 +\bar \eta \pa_\mu    \xi^\mu \Big]  
   \nn\\
  \sim -\int   \d t\ \d ^{d-1}x \Big[
     {{     
     \bar \xi^ i   \Big  (   \bgij    (  \bar \Delta +\A \pa_0 ) +   \frac{d-3}{d-1}  \pa_i\pa_j    \Big  )    \xi^i
      }}
      -  \eta '      \pa_i    \bar\xi^i    +
 \bar \eta  \pa_i \xi^j
      +{ \bar \xi^0}'   \pa_0\xi^0\Big] .
  \eea 
    with  $\eta'    =  \eta    + \A \xi^0$ and  $  { \bar \xi^0}' =\bar \xi^0+\bar\eta$.
    All    propagators  between       $(\xi^i,\xi^0,\eta)$ and
 $(\bar \xi^i,\bar \xi^0,\bar \eta)$ are   invertible.   
The gauge fixing that we have constructed is thus well defined perturbatively.  In the absence of gravitational anomaly,  the Ward identities   that control all interactions are  enforced by the BRST  invariance and{ ensure}  the  gauge invariance  of the observables and the unitarity order by order in perturbation theory. Eventually the class of   $\alpha$-unimodular  gauges     introduced  by the above discussion combine both its physicality and   narrow contact with the classical properties of the theory  as well as its consistency. 

   \section{Conclusion}

Given  the  classical  gravity action 
 $
I[ \gij, N, {\N }]   =
\int^{t_F, g_{ij}^F}_{t_I, g_{ij}^I}
 \d^d x \  L ^{Einstein}   (\gij, N, \N)
$,
the    expression   (\ref{try})   of the gravity observables  has been       made explicit in Eq.~(\ref{Egfl}). 
The    path integral  is  over all possible  internal  metrics  of leaf trajectories between two    boundary  leafs  at times $t_I$ and $t_F$.  The  class of $\A$-unimodular gauges determines a positive kinetic energy term    $ D_0 \bar g _{ij}   D_0 \bar g ^{ij} $  in a path integral formulation and   the    conformal  factor   decouples.  We have discussed the differences between  both choices 
    $\A\neq 0$ or $\A= 0$.    The  propagation  of the shift  vector~$N^i$~is    formally  analogous to     that   of the temporal component of a gauge field   in   a regularized   Coulomb gauge.
  
     The case $\A\neq 0$  gives      unambiguous  field propagators    both in ${\cal M}_d$ and in each leaf $\Sig _{d-1}$.
      The comparison with  the gauge fixing  
      in  the  regularized Coulomb gauge  of Yang--Mills fields{     enlightens  some   aspects of the gravity compactification       into lower  dimensional  gravity   and Yang--Mills theories, with 
          shift components  $g_{M0}$   identified as    time components  $A_{0}$ of a  gauge field with  spatial components     $A_k$ made  of   corresponding  $g_{Mk}$  components}.

      Eq.(\ref{Egfl})  is  a    BRST invariant path integral definition  that involves   the $\frac{d(d-1)}{2} -1$ off shell components $\bgij$ of the unimodular parts of    the  $d-1$ dimensional internal metric  $\gij$ as the    genuinely propagating   components  of the  quantum gravity field $\gmn$. It    identifies   ab initio  the 
   $ \frac{d(d-3)}{2}$ physical quanta of  gravity as the physical degrees of freedom of the quantum theory. 
          The  BRST invariance  of the unimodular gauge fixing  implies     Ward identities  for the observables $<{\cal O}  (\bgij)  >  $'s.
  They must  ensure that   a  physical quantum propagation  with unitary properties occurs  only for   the unimodular parts of the  internal  leaf metrics  (for  $d\geq 4$). The unimodular leaf gauge  can be viewed as a physical choice because   
    the classical limit of the corresponding   quantum theory  is then     automatically   compatible with the classical   properties  found in  \cite{ York}.
 Unphysical fields propagate in this gauge, but, owing to the BRST symmetry, their contributions compensate each other  in  the physical amplitudes. The  validity of this reasoning at the non perturbative level is  a question worth further~investigation.
 
   One must note       that  the  lapse function  has been gauge fixed   in a BRST invariant way with  $N(x^i,t) =1 $,    but  it  can also  be  fixed more generally  to any possible fixed positive function $N(x^i,t) = f(x^i,t)$, still in a BRST invariant way.  This generalization might be useful to  perhaps solve more  conveniently  certain problems.    Let us conclude by emphasising that   the physical   equivalence between Weyl related  internal metrics of each  leaf is quite subtle   and goes beyond   topics of  the genuine covariance under reparametrization symmetry. One might  be tempted  to  consider this  property     as  a defining  concept of the gauge invariance of general relativity.

    {
 \bluecor
A rather technical   and  relevant  comment  is that   practical calculations of quantum gravity   often use  the  background field method.   Both  subjects    are  not  the subject of this paper, but  they are very  interesting.  One  might thus attempt to  further  investigate the    possibility of defining  the background field   method within  the ADM   formulation, 
maybe  with the help of  some of  the advances presented  in this work. In fact,  since the ADM fields  
$N, N^i, g_{ij}$   undergo the whole repametrization transformations  in a one to one correspondence with those  of  $  g_{\mu\nu}$,   using them within the   background field formalism    seems   
a priori doable, even if they get split  in 
Weyl invariant and non Weyl invariant components,  as it is done systematically in the unimodular gauge. 
On the other hand, one must solve the  non trivial issue of   wether or not one has the necessity   of  introducing    background fields for both  the lapse and shift fields,  which  have no relevant physical propagation,  as well as for the conformal factor. The latter  questions are non trivial and deserve further studies. }

      \vskip 0,1cm
   \noindent {\bf{Acknowledgements:}} 
        I am most grateful to John Iliopoulos for  many discussions on the subject  of this work.  
   
  
\appendix
\section{Regularized Coulomb gauge}\label{App1}
\def\A{\alpha}
{  
{
\subsection{  Yang--Mills}
Consider  the abelian curvature $F_{ij} =\pa_i A_j-\pa_jA _i$  and  $F_{0i} =\pa_0 A_i-\pa_i  A_0$ of the gauge field $A_\mu=(A_i, A_0)$.  $F_{\mu\nu}^2 =  F_{ij}^2  -  F_{0i}^2$ is gauge invariant 
under $sA_\mu=\pa_\mu c$. 
With the gauge  choice 
$\A  A_0 =  \pa_iA_i $, one has 
\bea \label {coul} {F_{\mu\nu}^2 } _{|_{A_0 =\A ^{-1}  \pa_iA_i }}
&=& 
  -   A_i \Box  A_i   + ( \pa_i A_i )^2       +\A^{-2}   (\pa_i A_i) \Delta  (\pa_i A_i) +  \  boundary \ term
  \nn\\
&=&   -  A_i (\delta_{ij} -\frac { \pa_i\pa_j}{\Delta})   {\Box}  A_j  - \pa_i A_i  \frac   {\pa^2_0         -\A^{-2} \Delta^2} {  \Delta  }\pa_i A_i + 
\ 
boundary \ term
\eea
 where $\Delta=\pa^i\pa_i$. The propagator of  $A_i$ stemming from this gauge fixing of $F_{\mu\nu}^2$ is 
\bea \label{coulomb}
<A_i,A_j>  =- (\delta_{ij} -\frac { \pa_i\pa_j}{\Delta}) \frac{1} {\Box} +\frac { \pa_i\pa_j}{\Delta}   \frac{1}{\pa_0^2 -\A^{-2} \Delta^2}
        \eea
        A ghost  propagation  must occur to compensate for that of  unphysical modes implied by the  longitudinal part~in~(\ref{coulomb}).   It is defined    by  the  Faddeev--Popov ghost propagator one gets  from   the enforcement of a BRST invariance, by  adding the  term   $\bar c s(\A A_0 +\pa_i A^i)=     \bar c (\A\pa_0 +\Delta)c$    to the action (\ref{coul}). This ghost propagator is     
        \bea \label{coulombg}
<\bar c, c>  =   \frac{1}{\A \pa_0 + \Delta}.
        \eea
            Beyond     technical difficulties, the regularised Coulomb gauge is well-defined from (\ref{coulomb}), and, perturbatively, can be shown to yield the same physical results as   the Lorentz gauge $\pa_\mu A_\mu=0$ (or, more subtly, $  \A\pa_0 A_0 +\pa_i A^i=0$)~\cite{lbdz}.
        
       The  propagator (\ref{coulomb})  is to be compared with that of the  pure Coulomb  gauge  condition $ \pa_iA_i =0$ for which
     \bea \label{coulombpure} {F_{\mu\nu}^2 }  _{|_{   \pa_iA_i =0}}
      =   - A_i \Box  A_i   + A_0\Delta A_0+
\ {\rm
boundary \ term}.\eea 
The  gauge field   propagator implied by     (\ref{coulombpure})  suffers from the pure time dependent zero modes of $A_0$. The limiting case  $\A= 0$ is generally ambiguous because of contact terms, except for  questions where one can consistently set $A_0=0$ everywhere   through the computations.

  For Yang--Mills,   the possibility of practical perturbative computations comes from   the following    ``improved"    successions  of gauge functions: the ``physical''  but 
ambiguous gauge $\pa_i A_i =0\to  $   the   non ambiguous  gauge $\pa_i A_i =\A A_0,  (\A\neq0),  \to  $  the non ambiguous    renormalisable but  not so easily usable   gauge  $ \pa_i A_i  =\A \pa_0A_0 \to $ the non ambiguous  and  more easily renormalisable  gauge    with  $\A=1$, $\pa_i A_i = \pa_0A_0$, as  it is    Lorentz invariant.
     \def\g{\gamma}  
    \subsection{  A simpler but less physical gauge--fixing for gravity inspired from the Yang-Mills Coulomb gauge}
    Consider the following gravity simpler gauge-fixing in the ADM parametrisation,  where no separation between the conformal and non conformal part of both
 $\gmn$ and $\gij$  is made. The BRST   invariant action  for the gauge functions   $N^i   +\A   \pa_j g^{ij}=0$ and   $N-\g   {\dg}=0$, where   $\alpha \neq0 $ $\gamma\neq0 $ are a pair of  gauge parameters, is 
 \bea
  & & \quad \quad \quad \quad \quad \quad  \quad \quad  \int   \d^ d   x    \Big [   \sqrt {-g} R^{d}  +
 s(  \bar \xi ^i   (g_{ij}  
  \pa_k  g^{jk} 
   +\A 
   N_i) ) + s( \bar \xi ^0 (N-\g\dg\ ))  \Big ] \nn\\&=&
   \int   \d^ d   x   \Big [   \sqrt {-g} R^{d}  +
 b _i    ( \pa_j  g^{ji}  +\A N^i)  + b^0 (N- \g \dg)
 - \bar \xi ^is   (g_{ij } \pa_k  g^{jk}  +\A sN_i)  - \bar \xi ^0 (sN-\g s\dg\ ) \Big]
 \nn\\  &\sim & 
 \int 
 \d^{d-1} x \d t  
 \Big [
   {\frac { 1 }{\color{black} \g}   g  ^{ijkl }    D_0   g _{ij}   D_0   g _{kl}} { |_{N^i   =-\A   \pa_j g^{ij}}}
   +     \g { \color{black}   {\det \gij} }R^{d-1} (   g _{ij} )
 - \bar \xi ^is   (g_{ij } \pa_k  g^{jk}  +\A sN_i)  - \bar \xi ^0 (sN -\g s\dg\ )
  \Big ]. \nn
 \eea
  For $\gamma\neq0$, the ghost part of this   BRST invariant  action has  an       invertible  quadratic approximation, namely, 
 $
    - \bar \xi ^is   (g_{ij } \pa_k  g^{jk}  +\A sN_i)  - \bar \xi ^0( sN  -\g  \dg \  )
     \sim 
   - \bar \xi^i     (\Delta +\A\pa_0) \delta _{ij} +  \pa_i\pa_j) \xi^j   
    - \A 
    \bar\xi^i  \pa_i \xi^0
    -\bar\xi^0 (\pa_0\xi^0 -\g \pa_i\xi^i)
    .
    $
   The  quartet (\ref{gg}) is unneeded  in  such     classes of  $\alpha,\gamma$ gauges.  
   In these   perfectly   admissible but non-unimodular      gauges,       
    the conformal  factor  propagates    with a   negative contribution to the kinetic energy   (\ref{dW}).  The reason for that is     the contribution of the     opposite sign of the trace part of the  metric  tensor $g^{ijkl}$ over the space of~$d-1$~dimensional Euclidean  metrics  when the internal leaf metric is not gauge fixed in an unimodular way.} The relevance of the Weyl invariance for defining the observables is not as explicit in the classes     of  $\alpha,\gamma$ gauges than it is  in the   unimodular gauges.

    }\begingroup\raggedright\endgroup


\begin{thebibliography}{}
\bibitem{Arnowitt:1962hi}
R.~L.~Arnowitt, S.~Deser and C.~W.~Misner,  {\it  The dynamics of general relativity},
{{Gen.~Rel.~Grav.}~{\bf 40}~(2008)~1997--2027}.
 
%
\bibitem{York}
J.~W. York, Jr., {\it  
Role of conformal three-geometry in the dynamics of gravitation},
{{Phys.~Rev.~Lett.}~{\bf 28}~(1972)~1082--1085}; 
{\it  Conformally invariant orthogonal decomposition of
symmetric tensors on Riemannian manifolds and the initial value problem
of general relativity},
{{ J. Math. Phys.} {\bf 14}~(1973) 456--464}.


\bibitem{Thomas199}
J.~M.~Thomas,  {\it  Conformal correspondence of Riemann spaces},
{{  Proc. Nat. Acad. Sci.} {\bf 11} (1925) 257--259}.

\bibitem{Thomas2}
J.~M.~Thomas,   {\it   Conformal invariants},
{{  Proc. Nat. Acad. Sci.} {\bf 12} (1926) 389--393}.

\bibitem{Thomas3}
O.~Veblen and J.~M. Thomas, {\it  
Projective invariants of affine geometry of paths},
{{Ann.~Math.}~{\bf 27}~(1926)~279--296}.



\bibitem
{Einstein}
 A.  Einstein, {\it {Die Grundlage der Allgemeinen Relativitatstheorie}}, Annalen der Physik,~{\bf 354}  769-822, 1916. 



\bibitem{bcw}
L.~Baulieu,  L.~Ciambelli, and S.~Wu,
 {\it Weyl Symmetry in Stochastic Quantum Gravity},
     Class.~Quant.~Grav.~{\bf 37}~(2020)~4,~[hep-th]1909.11478.

\bibitem{unimodular} L. Baulieu,
{\it Unimodular Gauge in Perturbative Gravity and Supergravity}, Phys.Lett.~{\bf B808}~(2020),~[hep-th]~2004.05950.





\bibitem{Henneaux}
M. Henneaux and C. Teitelboim,
{\it{ The Cosmological Constant As A Canonical Variable}}
Phys.Lett.  B~{\bf 143} (1984) 415-420;
{\it{ The Cosmological Constant and General Covariance}}
 Phys.Lett. B~{\bf 222} (1989) 195-199.

\bibitem{Alvarez1} 
E.  Alvarez, {\it {The Weight of Matter}}, arXiv:1204.6162.

 \bibitem{Padilla}
A. Padilla, I. D. Saltas,  {\it {A Note on Classical and Quantum Unimodular gravity}},
arXiv:1409.3573.


 \bibitem{Bufalo}
S. Upadhyay, M. Oksanen, R. Bufalo,  
{\it {BRST Quantization of Unimodular Gravity}},  Braz. J. Phys.~{\bf47} (2017),  
arXiv:1510.00188.

\bibitem{Alvarez2}

E. \'Alvarez, S. Gonz\'alez-Martin, C. P. Martin, {\it {A note on the Gauge Symmetries of Unimodular Gravity,}}
arXiv:1604.07263.

\bibitem{Oda1}
  I. Oda,  {\it Fake Conformal Symmetry in Unimodular Gravity}, Phys. Rev. D{\bf 94} {044032} (2016), arXiv:1606.01571.

\bibitem{Oda2}
I. Oda, {\it {Classical Weyl Transverse Gravity,}}
arXiv:1610.05441.



\bibitem{Percacci}
R. de Leon Ardon, N. Ohta, R. Percacci, {\it {The Path Integral of Unimodular Gravity, }}
arXiv:1710.02457.

 
 
 

  
 
 \bibitem{Nagy}
S. Nagy, A. Padilla, I. Zavala, 
{\it {The Super-Stuckelberg procedure and dS in Pure Supergravity, }}
arXiv:1910.14349.

 
     
\bibitem{Martin}
J. Anero, C. P. Martin, R. Santos-Garcia,
{\it Off-shell unimodular N=1, d=4 supergravity},
JHEP (2020) 145,
arXiv:1911.04160,  {\it A note on unimodular N=1,d=4 AdS supergravity}, arXiv:2001.05365.



\bibitem{Vikman}
P. Jirousek, A. Vikman,
{\it New Weyl-invariant vector-tensor theory for the cosmological constant}
JCAP 04 (2019) 00,
e-Print: 1811.09547 [gr-qc]
https://inspirehep.net/literature/1704721; 
K. Hammer, P. Jirousek, A. Vikman
{\it } Axionic cosmological constant
e-Print: 2001.03169 [gr-qc],
https://inspirehep.net/literature/1774715.
 


\bibitem{lbdz}
L.~Baulieu and D.~Zwanziger
{\it Renormalizable noncovariant gauges and Coulomb gauge limit},
  Nucl.~Phys.~{cal B 548 }(1999) 527-562, 
  hep-th/9807024 [hep-th].
 
\bibitem{Baulieu:1986hw}
L.~Baulieu, C.~Becchi and R.~Stora,
{\it On the Covariant quantization of the free bosonic string},
{{Phys.~Lett.}~{\bf B180}~(1986) 55--60};
L.~Baulieu and M.~P. Bellon,
{\it  Beltrami parametrization and string theory},
{{Phys.~Lett.}~{\bf B196}~(1987)~142--150};
L.~Baulieu, M.~P. Bellon and R.~Grimm,
{\it  Left-right asymmetric conformal anomalies},
{{  Phys.\ Lett.} {\bf B228} (1989) 325--331}.









\bibitem{Kiefer:2017nmo}
C.~Kiefer and B.~Nikolic, {\it  
Conformal and Weyl-Einstein gravity: classical geometrodynamics},
{{  Phys.~Rev.}~{\bf D95} (2017) 084018}, [gr-qc]   1702.04973.

\bibitem{Gourgoulhon:2005ng}
E.~Gourgoulhon and J.~L.~Jaramillo, {\it  
A 3+1 perspective on null hypersurfaces and isolated horizons},
{{  Phys. Rep.} {\bf 423} (2006) 159--294},  [gr-qc]   0503113.


E.~Gourgoulhon,  {\it  3+1 formalism and bases of numerical relativity}, [gr-qc]   0703035.
 
 \bibitem{Gourgoulhon:blau}
M.~Blau,  {\it  Lectures Notes in General Relativity},
http://www.blau.itp.unibe.ch/newlecturesGR.pdf.

\bibitem{PhysRev.160.1113}
B.~S.~DeWitt, {\it   Quantum theory of gravity: the canonical theory},
{{  Phys. Rev.} {\bf 160} (1967) 1113--1148}.

\bibitem{Giulini:1994dx}
D.~Giulini and C.~Kiefer, {\it 
Wheeler-DeWitt metric and the attractivity of gravity,}
{{Phys.~Lett.}~{\bf A193}~(1994)~21--24}, [gr-qc]   9405040.
 
\bibitem {bb}
L. Baulieu and M. Bellon
{\it   p-Forms and Supergravity: Gauge Symmetries in Curved Space}, Nucl.Phys.~{\bf B~266 }(1986) 75-124.


\end{thebibliography}
\end{document}